\documentclass[prl,twocolumn,showpacs,amsmath,amssymb,superscriptaddress]{revtex4}
\usepackage[dvips]{graphicx}
\usepackage{graphics}
\usepackage{dcolumn}
\usepackage{bm}

\begin{document}
\title{Exact   diagonalization  study  of
domain structure in integer filling factor quantum Hall ferromagnets
}        
\author{E.H.  Rezayi}
\affiliation{Department of Physics, California State University, Los
Angeles,   CA   90032}   
\author{T. Jungwirth}
\affiliation{Institute   of   Physics   ASCR,
Cukrovarnick\'a 10,  162 53  Praha 6, Czech  Republic} 
\affiliation{
University  of  Texas at  Austin,  Physics Department,  1
University Station C1600,  Austin TX 78712-0264 
}  
\author{
A.H. MacDonald}
\affiliation{
University  of  Texas at  Austin,  Physics Department,  1
University Station C1600,  Austin TX 78712-0264 
}  
\author{
F.D.M. Haldane}
\affiliation{Department   of  Physics,  Princeton  University,
Princeton, NJ 08544-0708} \date{\today} 
\begin{abstract}

Opposite spin Landau levels in a quantum well 
can be brought into coincidence by tilting the magnetic field 
away from normal orientation.  We demonstrate that the  
magnetotransport anomaly at integer filling factors that was recently 
discovered by Pan {\it et al} is due to such a coincidence.  
By performing exact diagonalization calculations using microscopically
evaluated effective electron-electron interactions, we are able to 
establish that the electronic ground state at coincidence 
is an Ising quantum Hall ferromagnet and that the low energy
excitations correspond to the formation of a domain wall.

\end{abstract}


\maketitle

A  magnetic  filed component  parallel  to the  plane  of  a quasi  2D
electron  gas  can trigger  transitions  between correlated  many-body
states   in  the   quantum  Hall   regime.   For   example,  transport
anisotropies  induced   \cite{panprl99,lillyprl99}  by  weak  in-plane
fields at  Landau level filling  factors $\nu=5/2$ and  $\nu=7/2$ have
been   described   theoretically   \cite{rezayiprl00},   using   exact
diagonalization techniques,  as arising  from a transition  between an
incompressible  paired quantum  Hall state  and a  compressible stripe
phase. In a separate tilted-field magnetotransport experiment Pan {\em
et  al.}\cite{panprl00}  discovered,   at  $\nu=9/2$  and  $\nu=11/2$,
transitions, at critical tilt angles, between isotropic Fermi liquid
states and anisotropic stripe states.  An understanding of transitions
like these  that occur at  large tilt angles  can be achieved  only by
accounting for the large  changes in Landau level wavefunctions caused
by  the tilts.   The transitions  studied by  Pan {\em  et  al.}  were
explained  \cite{panprl00}  by  combining  many-body  RPA/Hartree-Fock
theory   with  self-consistent   local-spin-density-functional  (LSDA)
descriptions  of  one-particle   states  in  the  experimental  sample
geometry \cite{jungwirthprb99}.  Crudely  described, the effect of the
in-plane field in these experiments is to produce a crossing between a
valence Landau level  at small tilts, which is  the lowest quantum well
kinetic  energy eigenstate  of the  second subband  ($N=0,i=2$),  and a
Landau level that is a higher quantum kinetic energy eigenstate of the
first  subband  ($N=2,i=1$).   Stripe   states  occur  only  when  the
half-filled valence Landau level has dominant $N=2$ character.

In    this   paper   we    address   the    magnetotransport   anomaly
\cite{panprb01,remzeitlerprl01}  observed  in  the same  quantum  well
sample \cite{panprl00} at {\em  integer} filling factors for which the
stripe  states  that are  believed  to  be  responsible for  transport
anisotropy  in half-filled Landau  levels have  not yet  been observed
\cite{demler0110126}.  By performing  detailed quantum well electronic
structure calculations we are  able to establish that the experimental
anomalies occur at the point where the ($N=2,i=1$) and the ($N=0,i=2$)
Landau levels  cross.  We find that  the ground state at  the point of
coincidence   is  always   a  uniform   density  Ising   quantum  Hall
ferromagnet\cite{jungwirthprb01},  not  a  stripe state.   We  discuss
possible    sources   of   dissipation    and   anisotropy    in   the
finite-temperature transport  in these states, in  particular the role
of the domain wall excitations \cite{jungwirthprb01}.

\begin{figure}[h]


\includegraphics[width=3.3in]{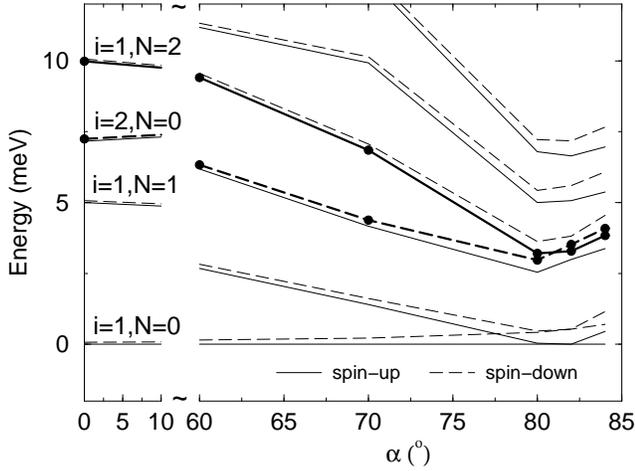} 

\vspace{0.5cm}

\caption{Self-consistent LSDA Landau level diagram plotted as a function
of the magnetic field tilt angle. The numerical data are obtained
for $\nu=6$ and sample parameters of Ref.~\protect\cite{panprb01} for
tilt angles indicated by filled circles.
(Lines are plotted to guide the eye.) Solid
(dashed) lines correspond to spin-up (spin-down) levels. The thick
lines simultaneously approach the chemical potential at $\alpha=81^{\circ}$.
        }
\label{lls}
\end{figure}

The sample  studied by  Pan {\em et  al.}  \cite{panprb01}  contains a
35~nm  wide GaAs  quantum  well with  2D  electron density  4.2$\times
10^{11}$~cm$^{-2}$. Due  to the relatively large thickness  of the quasi
2D system and the high  electronic density, the orbital effects of the
in-plane magnetic field, $B_{\parallel}$, play an especially important
role in determining many-body ground and excited state properties.  In
Fig.~\ref{lls}  we  plot single-particle  Landau  level  spectra as  a
function     of      the     magnetic     field      tilt     $\alpha$
($B_{\parallel}=B\sin\alpha$),  calculated  using the  self-consistent
LSDA at  filling factor $\nu=6$.  At $\alpha=0$,  the highest occupied
energy  level is  the  spin-down,  $N=0$ Landau  level  of the  second
subband   ($i=2$),   and  the   lowest   empty   states  are   spin-up
quasiparticles with  indices $N=2$  and $i=1$.  For  non-zero in-plane
fields, $N$ and $i$ are not  good quantum numbers and the orbital part
of  the eigenfunction is  described by  a single  index $n$.   The gap
between the $n=3$ level (emerging  from the $N=0$, $i=2$ Landau level)
and  the $n=4$  level (emerging  from the  $N=2$, $i=1$  Landau level)
decreases  with  increasing   $B_{\parallel}$,  eventually  leading  to
coincidence    at    a   high    tilt    angle.     Note   that    the
$B_{\parallel}$-enhancement of  the Zeeman gap  plays a minor  role in
bringing  the   two  levels  into   the  coincidence.   In   the  LSDA,
self-consistency  pins the  two levels  close together  over  a finite
range of $\alpha$, hinting at  the important role of correlations that
are not  captured by what is  effectively a mean-field  approximation for
this system.  As seen in  Fig.~\ref{lls}, the two levels remain nearly
degenerate  even  at  tilt  angles  as  large  as  84$^{\circ}$.   The
observation supports our contention  of the close relation between the
quantum  Hall ferromaget  physics and  the measured  transport anomaly
which, after  an abrupt onset around tilt  angle 82$^{\circ}$, remains
nearly   unchanged   up   to   the   highest   measured   tilt   angle
$\alpha=84.4^{\circ}$ \cite{panprb01}.

Within Hartree-Fock theory, the ground  state of a quantum Hall system
at integer filling  factors with two levels near  the Fermi energy can
be    described   by   a    Landau   energy    function   $e_{HF}(\hat
m)=-bm_z-U_{z,z}m_z^2$ \cite{jungwirthprb01}.  In our case, pseudospin
$m_z=-1$ corresponds to full occupation of quasiparticle states in the
$n=3$ spin-down  level and $m_z=+1$ corresponds to  full occupation of
the $n=4$ spin-up  Landau level.  The sign and  magnitude of $U_{z,z}$
can be  evaluated from the  microscopic wavefunctions of  the crossing
Landau levels.  \cite{jungwirthprb01} We  find that $U_{z,z}>0$ in the
present case, implying that the quantum Hall ferromagnet has easy-axis
(Ising)   anisotropy   \cite{jungwirthprb01}   in   the   Hartree-Fock
approximation.   Numerical  exact  diagonalization  studies  discussed
below confirm  that near the  level crossing ($b\approx0$)  the ground
state remains fully pseudospin polarized and uniform.

\begin{figure}[h]

\vspace{-1cm}

\includegraphics[width=3.3in]{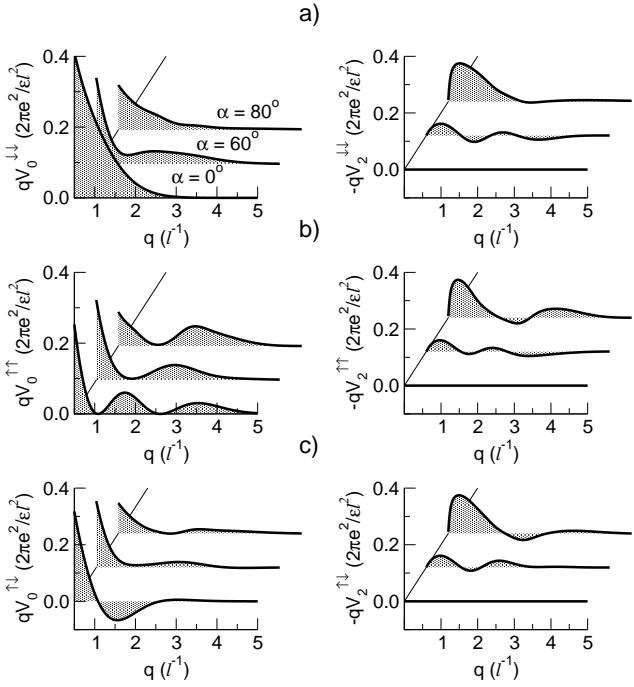} 

\vspace{0.5cm}

\caption{
Isotropic term (left panels) and the anisotropy
coefficient (right panels) of the effective 2D Coulomb
interaction multiplied by the wave vector amplitude $q$ at 
different magnetic field tilt angles. a) Intra-level potentials
for the $n=3$ ($i=2$, $N=0$ for $\alpha=0$) spin-down
Landau level. b) Intra-level potentials
for the $n=4$ ($i=1$, $N=2$ for $\alpha=0$) spin-up
Landau level. c) Inter-level potentials.
}
\label{coulomb}
\end{figure}

In   an  effort  to   achieve  a   more  quantitative   and  confident
understanding of the collective  behavior of electrons in this quantum
Hall  system, we  used  the self-consistent  LSDA one-particle  states,
obtained  numerically   for  the  specific   $\alpha$,  and  structural
parameters   of   the  sample   to   construct  realistic   effective
interaction potentials.  For the  $\nu=6$ quantum Hall ferromagnet we
obtained  two  intra  and   one  inter  Landau  level  potentials  for
electronic states  in the crossing  levels. The influence of  the five
remote, fully occupied Landau levels was captured perturbatively using
the  RPA dielectric  function  \cite{jungwirthprb99}. The  pseudo-spin
dependent  effective   interaction  potentials  can   be  approximated
\cite{panprl00,jungwirthprb99}            by            $V^{s,s'}(\vec
q)=V^{s,s'}_0(q)+V^{s,s'}_2(q)\cos(2\phi)$         where         $s,s'
=\uparrow$~or~$\downarrow$  are  pseudospin labels,  $q$  is the  wave
vector magnitude,  and $\phi$ is the wave  vector orientation relative
to the in-plane field  direction.  At $B_{\parallel}=0$ the pseudospin
dependent isotropic  effective interactions $V^{s,s'}_0(q)$,  shown in
the left  panels   of  Figs.~\ref{coulomb}(a)-(c),  have   a  wave  vector
dependence  similar   to  those  of  the   effective  interactions  in
infinitely  narrow 2D layers.   Due to  in-plane field  induced mixing
between  electric and  magnetic levels  in  the quasi  2D system,  the
$q$-dependence  of $V^{s,s'}_0(q)$  changes  with increasing  magnetic
field tilt  angle and the effective  interactions becomes anisotropic,
as seen from the  non-zero $V^{s,s'}_2(q)$ coefficients plotted in the
right panels of Figs.~\ref{coulomb}(a)-(c).

The  many-body Hamiltonian  with the  LSDA/RPA  effective interaction
potentials was  diagonalized numerically to obtain  the exact spectrum
in  a finite  size  system of  up  to twelve  electrons  in the  torus
geometry\cite{haldaneQHbook}.   This  geometry   is  well  suited  for
detecting  broken translational symmetry  phases.  The  calculation of
the  low  lying  spectrum  (for  example  as a  function  of  the  2-D
wavevector  ${\bf K}$\cite{haldaneQHbook})  amounts  to a  ``numerical
crystallography''\cite{HRY} in these phases.   In the case of stripes,
the degeneracy $N_D$  of the ground state manifold  corresponds to the
number of electrons per stripe  and their ${\bf K}$-vectors form a 1-D
array in reciprocal space  with $\Delta K_{ij}=K_i-K_j=nQ$, where $n =
1,2,\ldots N_D-1$, $i,j=1,2\ldots,N_D$,  and $\lambda=2\pi/|Q|$ is the
wavelength  of the stripes.   Because of  this property,  stripe ground
states  are readily  identified in  finite-size  exact diagonalization
calculations.   In  a finite  size  system  with  $N_e$ electrons  and
periodic    boundary    conditions,    the    interaction    potential
$V^{s,s'}(q_x,q_y)$ is only required at the $N_\phi^2$ values of ${\bf
q}$, where $N_\phi$ is the number  of flux quanta.  We obtain these by
interpolating the LSDA/RPA potential which is calculated at a discrete
set of ${\bf  q}$ values that do not in general  coincide with the PBC
values.  We  carried out our  calculations for tilt  angles $\alpha=0,
20, 40,  60,$ and  $80$ degrees  in a rectangular  unit cell  while we
varied the aspect ratio.  As shown previously\cite{HRY}, this helps to
reveal any  potential broken symmetry states.  In  all calculations we
have  restricted the  Hilbert space  to states  within the  two Landau
levels of interest.  (Processes which change the Landau level index of
an  electron  can be  neglected  since  they  require reversal  of  an
electron  spin.)  The  eigenstate energies  reported on  below  are in
units    of    $e^2/4\pi\epsilon_0\ell_\perp$,   where    $\ell_\perp$
corresponds to  the magnetic  length associated with  the out-of-plane
component of the magnetic field $B_\perp$.

\begin{figure}[t]

\vspace*{2cm}

\includegraphics[width=3.3in]{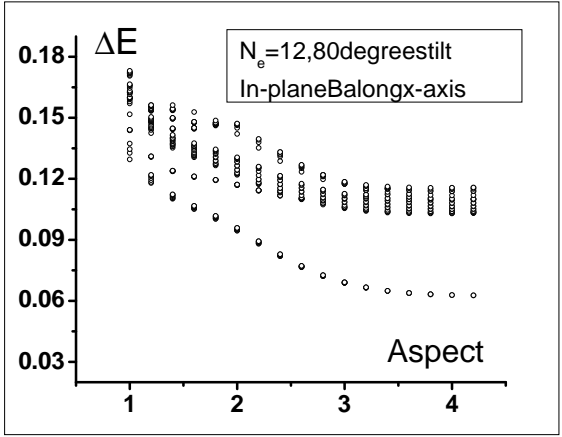} 

\vspace{-1.25cm}

\caption{Low-lying spectrum for 12 particles with 6 reversed spins
at total filling factor one as a function of aspect ratio at tilt
angle of 80 degrees. 
The energy of the Ising ($Z_2$) ground state has been subtracted out. The
lowest set of quasi-degenerate levels signify the formation of a 
domain wall between the two ground states.   
The domain wall begins to
develop for aspect ratios greater than 1.2.}
\label{spec80}
\end{figure}

For  {\em all}  tilt  angles $\alpha$  and  in {\em  every} unit  cell
geometry, we find that the ground state is a fully pseudospin polarized
uniform density  state.  For every  $\alpha$ we fixed  $\Delta_z$, the
Zeeman gap,  so that the  two polarized ground states  are degenerate,
{\em i.e.}  we  have set the effective pseudospin  field to zero.  The
entire  spectrum is  then $Z_2$  symmetric  as required  for an  Ising
ferromagnet.  In  the experiments  of Pan {\it  et al}  the degeneracy
occurs  for only  a single  $\alpha$  (about $82$ degrees).  By  fixing
$\Delta_z$  in  this   way  we  have  in  effect   studied  the  Ising
ferromagnet, if it were to occur, at a general tilt angle $\alpha$. We
find only quantitative differences  between different tilt angles; the
generic behavior  is independent of  any particular value  of $\alpha$
and is present even at zero  tilt.  While the ground state remains the
uniform density filled level, we  can identify low lying states over a
substantial range of aspect ratios  that correspond to the formation of
a domain  wall.  (These states have an  excitation energy proportional
to  the linear  size of  the system  and are  therefore  not low-lying
elementary excitations in the  thermodynamic limit.)  We did not find
any  stripe order, either  in the  ground  state or  in the  low-lying
excitations.   Fig.~\ref{spec80} shows how  the low  lying excitations
for a  tilt angle  of 80 degrees  change as  a function of  the aspect
ratio for a 12 electron system  with 6 reversed spins.  Here, the long
direction is along the y-axis and  the domain wall is always along the
x-axis    (short   direction). 
Fig.~\ref{spec80}  corresponds to an  in-plane field  along the  x
direction.     The    lowest    lying    manifold    has    degeneracy
$N_D=12=N_e=N_\phi$ with  wavevectors $(0,2m\pi/L)$, where  $L$ is the
length of the long side  and $m=0,1,2,\ldots N_e-1$.  This signifies a
phase separated state  with a domain wall along the  short side of the
unit cell\cite{footnote}.  

\begin{figure}[h]

\includegraphics[width=3.3in]{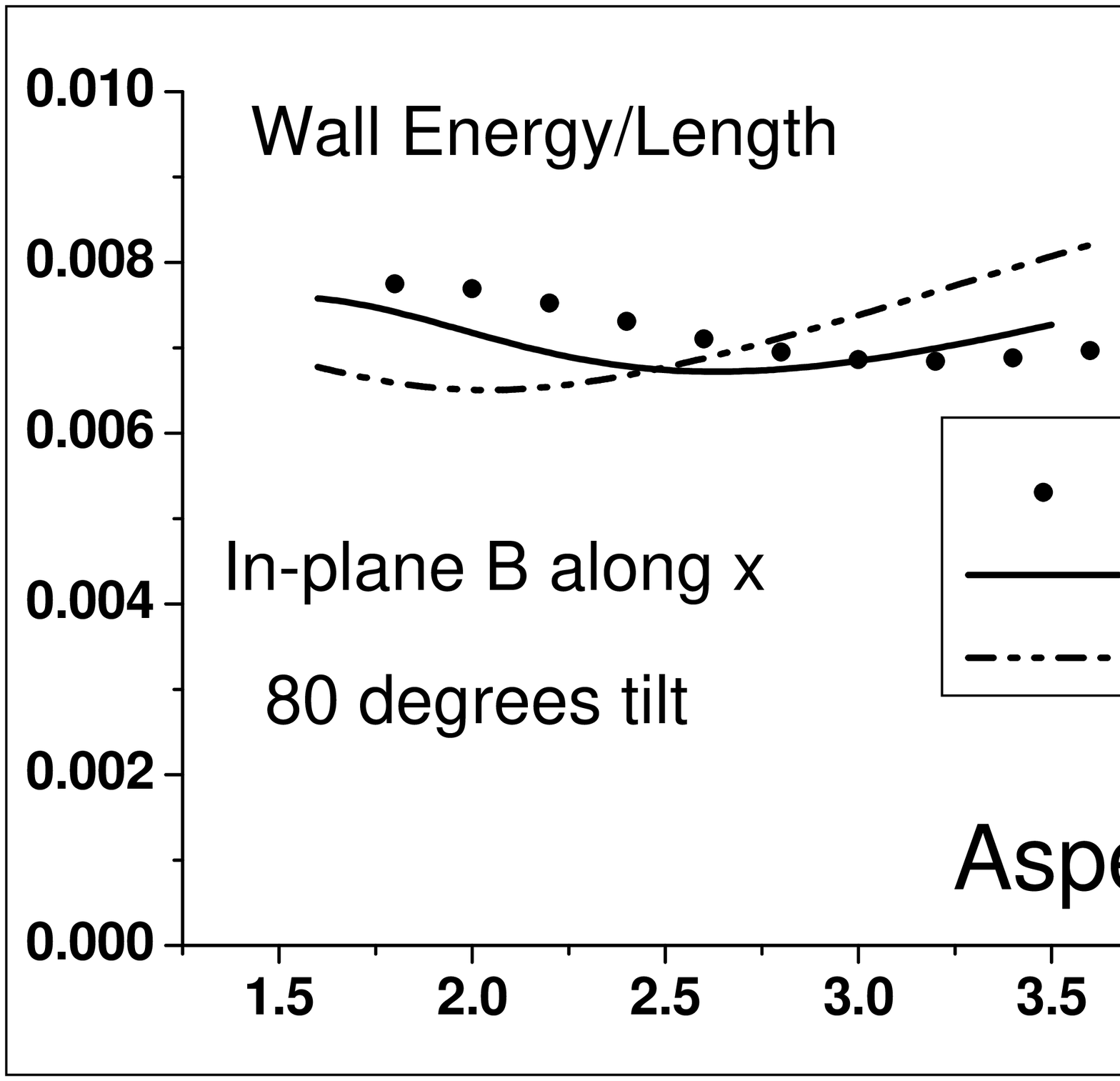} 

\vspace{-1.5cm}

\caption{The energy  of the domain wall  per unit length  for the same
system  as  in Fig.~\ref{spec80}.   Here  $B_\parallel$  is along  the
x-axis.  We  have only  included aspect ratios  for which the  wall is
well formed (see Fig.~\ref{spec80}).}
\label{wall80a}
\end{figure}

The domain  wall energy per unit length for
$N_e=8,  10$, and  12 electrons  is shown  in  Fig.~\ref{wall80a} 
which corresponds  to $B_\parallel$  applied  along the
x-axis and the magnetic field tilt angle of 80 degrees.  
For  other tilt  angles we  obtain  similar results,
except  that  the  energy  per  unit  length of  the  domain  wall  is
consistently reduced as the tilt angle is increased. Our domain wall
energy calculations
also confirm that the in-plane component of the magnetic field induces
an anisotropy in the domain structure of this Ising ferromagnet. Due
to finite size effects, however, we were unable to conclusively 
establish whether the parallel orientation or perpendicular orientation of
the domain wall is energetically more favorable for the studied quantum
Hall system. This issue can be addressed separately, e.g. 
within a self-consistent
Hartree-Fock theory, now that  the Ising quantum Hall ferromagnet nature
of the many-body system has been established by our exact diagonalization
study. This calculation, however, is beyond the scope of this paper.

In  concluding we  discuss  the relationship  between our  theoretical
findings  for the  one-particle and  many-body energy  spectra  and the
measured  \cite{panprb01} transport  anomalies.  In  a  previous study
\cite{jungwirthprl01},  we  proposed  that   enhanced  dissipation
would occur in  Ising quantum Hall ferromagnets as  a result of charge
diffusion along  domain walls.  If this  transport mechanism dominated
in the sample studied by Pan  {\em et al.}  \cite{panprb01}, a peak in
the longitudinal  resistivity would occur when the  system breaks into
domains with  opposite pseudospin orientations, i.e.,  near the Landau
level coincidence  \cite{jungwirthprl01}.  Our finding  that the $n=3$
spin-down and  $n=4$ spin-up Landau  levels are nearly  degenerate for
tilt  angles exceeding  $\alpha>81^{\circ}$,  is therefore  consistent
with this  interpretation of the experiments.  The  dissipation of the
quasiparticle current in a system with an anisotropic domain structure
could  provide a mechanism  for the  measured transport  anisotropy in
this  picture.  In the measured sample, the  in-plane field direction
is the  high resistivity direction  \cite{panprb01} so, intuitively,
the  domain  walls are oriented perpendicular to the in-plane field
direction.
Measurements of the temperature dependence of the transport anomaly in
samples with  different in-plane magnetic  field orientations relative
to the  crystal axes  and corresponding microscopic calculations, mentioned
in the previous paragraph, may
provide  useful information  on  how the  anisotropy  of tilted  field
quantum  Hall  ferromagnets  is  expressed  in  transport  experiments
\cite{chalkeretal} and on the combined effect of the in-plane magnetic
field and disorder on the preferred orientation of the domain walls.

\begin{acknowledgments}
We acknowledge useful discussions with Wei Pan and Horst Stormer, EHR 
would like to thank A. Davalos for help in exact diagonalization studies. The
work was supported by the  Welch Foundation, by DOE under contract
DE-FG03-02ER45981 (EHR), by the Grant Agency of the Czech Republic under
grant 202/01/0754, and by NSF DMR-0213706 (FDMH).
\end{acknowledgments}

\begin{figure}[h]
\includegraphics[width=3.3in]{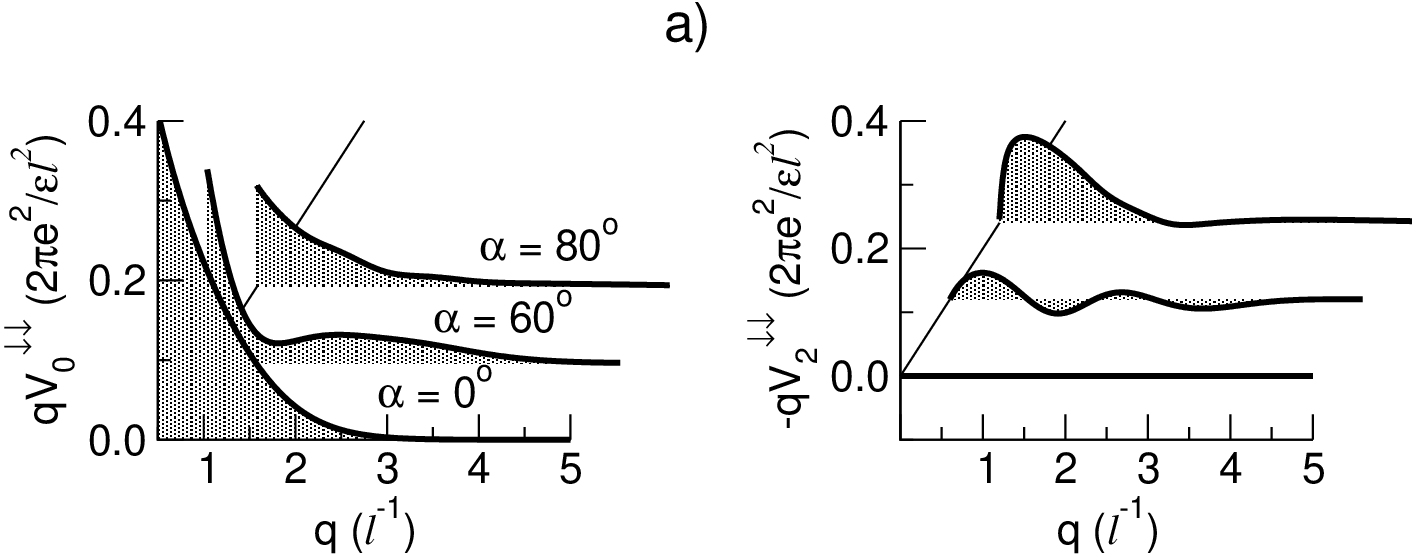} 


\end{figure}

\begin{figure}[h]

\vspace{-1cm}

\includegraphics[width=3.3in]{coul44_scr} 


\end{figure}

\end{document}